\documentclass[10pt]{article}
\usepackage{geometry}
\usepackage{booktabs}
\usepackage{titling}
\usepackage{enumitem}
\usepackage{xcolor}
\usepackage{hyperref}
\usepackage{graphicx}
\usepackage{tabularx}
\usepackage{amsmath}

\title{\textbf{Detecting AI Coding Agents in Open Source: A Validated Multi-Method Census of 180 Million Repositories}}
\author{Arsham Khosravani, Audris Mockus}
\date{}

\begin{document}

\maketitle

\begin{abstract}
Generative AI coding agents are rapidly entering the open-source software supply chain, yet their diverse and often invisible traces leave their true prevalence poorly understood. We introduce a multi-layered detection framework that integrates configuration-file scanning, commit-message analysis, author-identity pattern matching, and bot-signature lookup across the World of Code (WoC) infrastructure, spanning more than 180 million Git repositories.

Our four-type taxonomy shows that no single detection method captures more than a fraction of total AI activity. Multi-method detection identifies 850,157 Claude Code commits in the V2510 snapshot, of which bot-account lookup, the signal most adoption studies rely on, recovers only 28,154 (3.3\%): a $30\times$ relative-recall gap. Absolute recall is unknown; the union is a lower bound, so single-signal prevalence estimates are biased low by at least this factor. Every detection pattern is hand-validated (495 labels) with per-cell precision and Wilson confidence intervals.

Applying the framework across snapshots from December~2024 through April~2026, we find that commit-attributed agents collectively generate over 320,000 commits per month by V2604. Claude Code leads with 886,122 commits across 17,295 projects, followed by Jules with 215,804 commits. The Type~D configuration-file census further establishes Claude as the dominant silent agent in V2604, appearing in 21,078 projects (888,177 blob occurrences), a category entirely absent in V2412. Adoption timing is bimodal: projects either integrate agents at inception (born-with-AI) or after years of development. Comparing against an independent pull-request-based census (AIDev), we find the two channels capture nearly disjoint agent populations: Codex is the largest agent by pull requests but near-absent from commits, while Claude Code is the largest by commits but near-absent from pull requests, so a PR-based census misses 79\% of commit-detected Claude Code adopters and a commit-based census misses essentially all Codex adopters. No single detection channel is representative---and the channels differ in \emph{kind}, not just population: via AIDev's task labels, PR-deployed cloud agents (Codex, Cursor) surface as feature work while commit-deployed in-editor agents (Claude Code, OpenHands, Aider) surface as maintenance, so the observed agent work profile follows deployment and detection mode rather than the tool itself.
\end{abstract}

\section{Introduction}
\label{sec:intro}

Generative AI (GenAI) coding agents, tools that autonomously
generate, modify, or review source code, are increasingly embedded
in open-source software (OSS) development workflows. Tools such as
GitHub Copilot~\cite{peng2023impact}, Cursor~\cite{research2026composer}, Replit~\cite{replit2024agent}, and OpenHands~\cite{wang2024openhands} now participate in
activities ranging from autocompletion to autonomous pull-request
generation. Understanding how widely these agents are adopted and
what effect they have on development outcomes is important for at
least three reasons.

First, from a \emph{software engineering perspective}, quantifying
AI-assisted contributions enables researchers to study how GenAI
tools affect development effort, code quality, and project
sustainability~\cite{he2026speed}. Without reliable detection, any
impact study conflates human and machine contributions.

Second, from a \emph{software supply-chain} perspective,
AI-generated code introduces distinct risk profiles. It may embed
subtle vulnerabilities, propagate biased training-data patterns, or
create negative feedback loops where generated code degrades the
training corpus of future models~\cite{ji2024cybersecurity}.
Identifying AI-generated contributions is therefore a prerequisite
for supply-chain auditing.

Third, from an \emph{ecosystem measurement} perspective, the
explosive growth of coding agents documented in
2025~\cite{robbes2026agentic} demands baseline prevalence estimates
grounded in repository-level evidence rather than self-reported
surveys.

Despite this need, detection is non-trivial. Agents differ
dramatically in the traces they leave: some operate through
dedicated bot accounts, others embed signatures in commit messages,
and many are entirely ``silent,'' modifying code under the
developer's own identity with no distinguishing marker. A detection
strategy that relies on any single signal will systematically
undercount adoption, and prior censuses rely on exactly one signal
(pull requests or configuration files).

These gaps frame three research questions:
\begin{description}[leftmargin=2.4em,style=nextline]
  \item[RQ1 (prevalence).] How prevalent are AI coding agents across
    the OSS ecosystem, how much do single-signal methods miss, and
    do different detection channels see the same agents and the same
    kinds of work?
  \item[RQ2 (adoption timing).] When in a project's lifetime are
    agents adopted?
  \item[RQ3 (velocity association).] Does adoption coincide with
    changes in development velocity?
\end{description}
RQ1 demands the multi-method detection our taxonomy provides; RQ3 is
deliberately descriptive, establishing association rather than a controlled
causal estimate.

This paper makes three contributions that answer these questions:

\begin{enumerate}[label=(\arabic*)]
  \item A \textbf{detection taxonomy} (Types~A--D) that classifies
    AI coding agents by the traces they leave in version-control
    history, together with a multi-layered detection framework
    implemented on the World of Code infrastructure~\cite{ma2021world}.
  \item A \textbf{prevalence census} of twelve agents across
    180M+ repositories and three snapshots (December~2024 to
    April~2026), revealing that silent and distributed agents
    account for a substantial share of AI activity invisible to
    prior file-based and PR-based surveys. A direct comparison with
    an independent PR-based census (AIDev~\cite{li2026aidev}) shows
    the commit and PR channels capture nearly disjoint agent
    populations \emph{and} different kinds of work---feature generation
    in the PR channel, maintenance in the commit channel---so the observed
    work profile follows deployment and detection mode, not the tool, and no
    single-channel census is representative.
  \item A \textbf{hand-labeled validation} of every cell in the
    detection grid (495 labels) with per-cell precision and Wilson
    confidence intervals, a construct-validity analysis separating
    agent-authored from agent-trailered commits, and a documented
    audit trail of detector corrections.
\end{enumerate}

\section{Related Work}
\label{sec:related}

\paragraph{Adoption surveys.}
Robbes et al.~\cite{robbes2026agentic} provide the most
comprehensive adoption study to date, documenting the rapid
uptake of coding agents on GitHub during early 2025 by tracking
bot-authored pull requests. Their methodology is PR-centric,
capturing Type~A agents (centralized bot accounts) but missing
silent tools (Type~D) and distributed-attribution agents
(Type~C). Li et al.~\cite{li2026aidev} likewise build a large
PR-based census (AIDev, 932{,}791 agent-authored pull requests);
we compare against it directly (Section~\ref{sec:results}) and
show that the PR channel and our commit channel see nearly
disjoint agent populations. He et al.~\cite{he2026speed} use repository configuration files (\texttt{.cursorrules}) as an adoption proxy for Cursor, which corresponds to a single Type~D signal in our taxonomy. Our framework generalizes this approach. The same \texttt{.cursorrules} proxy recovers Cursor's 28{,}909 configuration projects (Table~\ref{tab:agents}), but taken alone it would report zero commit-level activity and would entirely miss Cursor's pull-request channel (32{,}941 PRs in AIDev; Table~\ref{tab:aidev}). Conversely, it is blind to commit-channel agents such as Claude Code. The single-signal proxy is thus correct but partial. We quantify exactly what the additional detection tiers contribute. A 2023 GitHub developer survey~\cite{github2023devsurvey} reports
that 92\% of surveyed developers use AI coding tools, but
self-reported surveys are susceptible to social-desirability bias
and cannot quantify actual commit-level contributions.
Each of these is a \emph{single}-channel measurement; None of these prior approaches establishes how much the other channels miss, or whether the different detection channels even observe the same agents and activity. This is precisely the representativeness gap that RQ1 targets. Adoption timing (RQ2) is likewise unmeasured at ecosystem scale.

\paragraph{Bot detection in software repositories.}
The problem of identifying non-human contributors in version
control predates GenAI. Dey et al.~\cite{dey2020detecting}
developed methods to detect and characterize bots that commit
code in OSS, identifying distinct patterns: burst timing,
formulaic messages, and limited file scope. Golzadeh et
al.~\cite{golzadeh2021ground} provide a ground-truth dataset
for bot detection in GitHub issue and PR comments. Wessel et
al.~\cite{wessel2018power} study how bots shape open-source
project interactions. These works focus on CI/CD bots and
dependency updaters (e.g., Dependabot, Greenkeeper), whose
behavioral signatures (predictable schedules, narrow file
scope, template messages) differ substantially from GenAI
agents that produce diverse, context-dependent code under
human-like identities. Our taxonomy extends bot detection to
this new class of generative agents.

\paragraph{AI-generated code detection.}
Beyond repository metadata, a growing literature addresses
detecting AI-generated code at the source level.
Shimonaka et al.~\cite{shimonaka2016detecting} pioneered
machine-learning detection of auto-generated code using AST
node distribution signatures. For LLM-generated code
specifically, Yang et al.~\cite{yang2023zeroshot} study detection via
token-level perplexity (lower for machine-generated text),
stylistic uniformity, and repeated higher-order $n$-gram
patterns, while Yin et al.~\cite{yin2025detecting} train
detectors adversarially to stay robust against subtle
modifications of LLM-generated code. Nguyen and Nadi~\cite{nguyen2022empirical}
characterize Copilot's code suggestions and their correctness.
Our detection framework operates at the version-control
metadata level rather than source-code content, making it
complementary to these content-based approaches and far more
scalable: we analyze 180M+ repositories without parsing code.

\paragraph{World of Code infrastructure.}
Ma et al.~\cite{ma2021world} describe the WoC
infrastructure, which indexes $>$180M Git repositories
into cross-referenced maps of authors, commits, files,
and projects. We use WoC's native hash maps for
O(1) author and commit lookups, its ClickHouse database
for commit-message analysis, and its file-to-project
maps for configuration-file scanning. The identity
resolution infrastructure~\cite{identity20} (aliasing
38M author IDs) is essential for accurate developer
counting in our outcome metrics. The deforking
methodology~\cite{forks20} prevents double-counting
of forked repositories in prevalence estimates.

\section{Methodology}
\label{sec:method}

\begin{figure*}[t]
  \centering
  \includegraphics[width=1.0\textwidth]{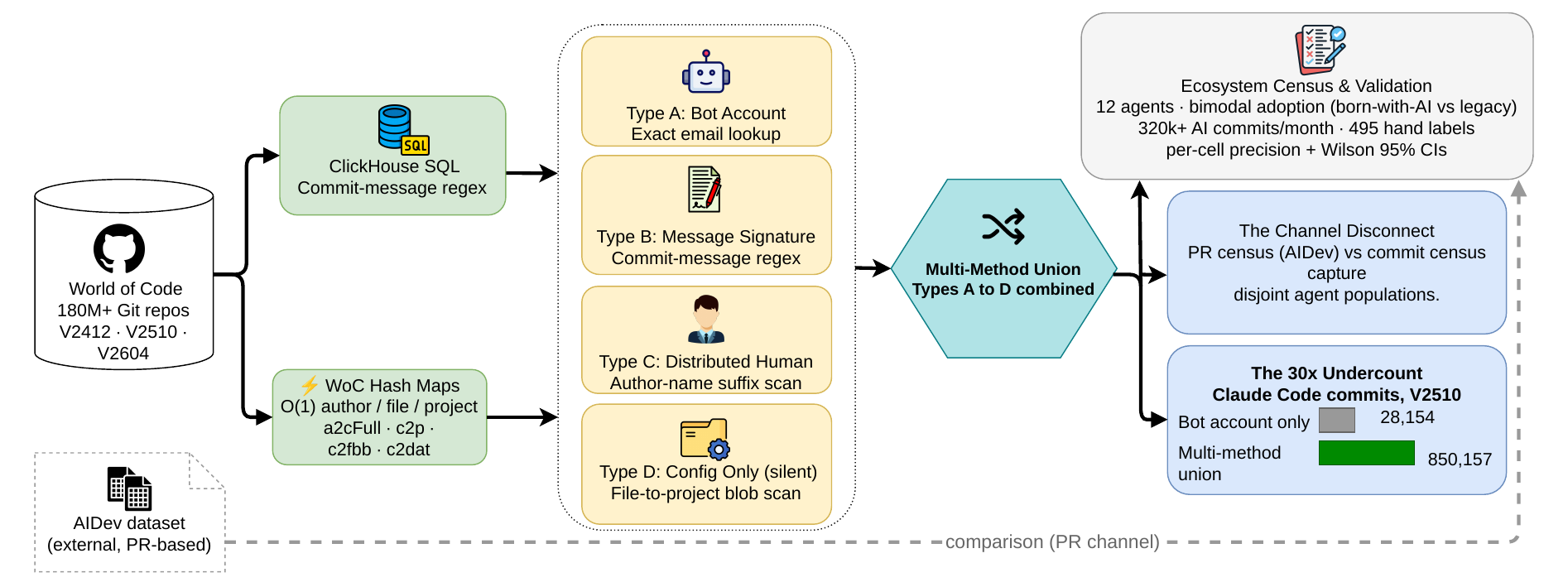}    
 \caption{Overview of the multi-method AI-agent detection framework.}
  \label{fig:overview}
\end{figure*}

\subsection{Data Source}

We analyze three World of Code (WoC) snapshots spanning December~2024 to April~2026: V2412 (December~2024), V2510 (October~2025), and V2604 (April~2026). Together, these snapshots index over 180 million Git repositories from GitHub, GitLab, Bitbucket, and other platforms. 

WoC structures software artifacts (authors, commits, blobs, files, and projects) into cross-referenced maps that support both SQL queries via ClickHouse and fast O(1) hash-map lookups. Prevalence results draw on all three snapshots, while impact analyses use the \texttt{commit\_v2510} dataset. V2412 and V2510 commit metadata are indexed in ClickHouse; V2604 is not yet ingested, so V2604 commit-level results derive from parallel scans of the \texttt{c2dat} commit-metadata and \texttt{c2fbb} file-blob flat-file maps (128 shards each), which carry the author, timestamp, and message fields the detectors require. Each successive snapshot also expands corpus coverage with newly created and newly discovered repositories, so cross-snapshot comparisons reflect both agent activity and corpus growth.

\subsection{Multi-Layered Detection Taxonomy and Framework}
\label{sec:taxonomy}

As illustrated in Figure~\ref{fig:overview}, our detection strategy classifies AI traces into four behavioral types, applying a complementary detection method to each:

\begin{description}[style=nextline]
  \item[Type~A,  Centralized Bot Account.]
    The agent commits under a single registered bot identity (name + email). We detect these via exact email-address lookups in the author map, providing $O(1)$ retrieval. 
    Examples: OpenHands~\cite{wang2024openhands} (7,972 commits), CodeRabbit~\cite{coderabbit2024} (297 commits).

  \item[Type~B,  Commit-Message Signature.]
    The agent embeds explicit text in commit messages. We query the WoC ClickHouse database using case-insensitive regular expressions to detect these signatures. 
    Example: Replit~\cite{replit2024agent}  (385,668 commits containing ``Generated by Replit'').

  \item[Type~C,  Distributed Human Attribution.]
    No central bot account exists; individual developers append a tool-specific suffix to their Git author name (e.g., \texttt{(aider)}). We identify these by scanning all 32 shards of the \texttt{a2cFullV2412} author map for the name pattern. 
    Example: Aider~\cite{gauthier2023aider}, 355 developers yielding 25,215 commits.

  \item[Type~D,  Configuration-File Only (Silent).]
    The agent leaves traces in fewer than 0.5\% of commits,
    making detection dependent on configuration-file presence.
    We query WoC file-to-project maps for agent-specific files
    (e.g., \texttt{.replit},
    \texttt{copilot-instructions.md},
    \texttt{CLAUDE.md}, \texttt{AGENTS.md}) using strict
    end-of-line anchoring to avoid false positives. This
    category is highly dynamic because agents frequently
    change configuration conventions, requiring periodic
    recalibration of detection rules. Example: GitHub
    Copilot (92{,}276 projects in V2412; 211{,}166 blob
    occurrences in V2604).
\end{description}

This taxonomy demonstrates that no single detection method
suffices: a complete census requires all four tiers operating in
combination.

\subsection{Author Aliasing and Bot Disambiguation}
\label{sec:identity}

Accurate counting requires resolving \emph{who} authored each commit. We rely
on WoC's identity-resolution infrastructure~\cite{identity20}, which links the
many \texttt{name <email>} strings a single developer uses---across machines,
and with email and casing variants or typos---into one resolved identity,
aliasing roughly 38M raw author IDs. The resolution connects author strings
that share strong identifying tokens (a common email local-part or a
distinctive normalized name) into developer clusters. This matters wherever we
report developers rather than commits: without it, one person's variant
spellings inflate contributor counts as several distinct developers.

Bot and agent identities are handled \emph{opposite} to human ones. A
centralized agent account (Type~A) is kept as its own identity and never merged
into a human alias cluster, so agent activity is not absorbed into a
developer's history (and, symmetrically, agents are excluded from human
contributor counts). Agents that instead commit under human identities are
caught not from the resolved identity but from the trace they leave: a
case-insensitive commit-message signature or co-authorship trailer (Type~B),
or a raw author-name suffix (Type~C). The Type~C tag (e.g.\ \texttt{(aider)})
is matched on the \emph{unresolved} author string precisely so that aliasing
cannot fold it away. The four detection primitives---exact case-insensitive
author-email lookup ($O(1)$ hash), commit-message regular expressions,
author-name-suffix scanning across all author-map shards, and
end-of-line-anchored configuration-file matching against the file-to-project
maps---each capture a different slice of the same agent, so we report their
\emph{multi-method union}; Finding~4 shows the union exceeds any single signal
by up to $30\times$. Deforking~\cite{forks20} is applied before all counts to
prevent fork copies from inflating prevalence.

\subsection{Non-GitHub Data Completeness}
\label{sec:completeness}

To ensure cross-forge coverage, we augmented the WoC gathering
pipeline using the Ecosyste.ms~\cite{ecosystems} API, which indexes repository
metadata across nearly 2,000 software forges. We identified 2,027
non-GitHub hosts and extracted 61,298 repository URLs from 1,500
previously untracked forges. The remaining 527 hosts (primarily
\texttt{gitlab.com} and \texttt{bitbucket.org}) are already covered
by WoC's native gathering methods.

\section{Results}
\label{sec:results}

\subsection{RQ1: How prevalent are AI coding agents across the OSS ecosystem?}

Table~\ref{tab:agents} summarizes the detection results for twelve
agents. Three findings stand out.

\begin{figure*}[t]
  \centering
  \includegraphics[width=0.7\textwidth]{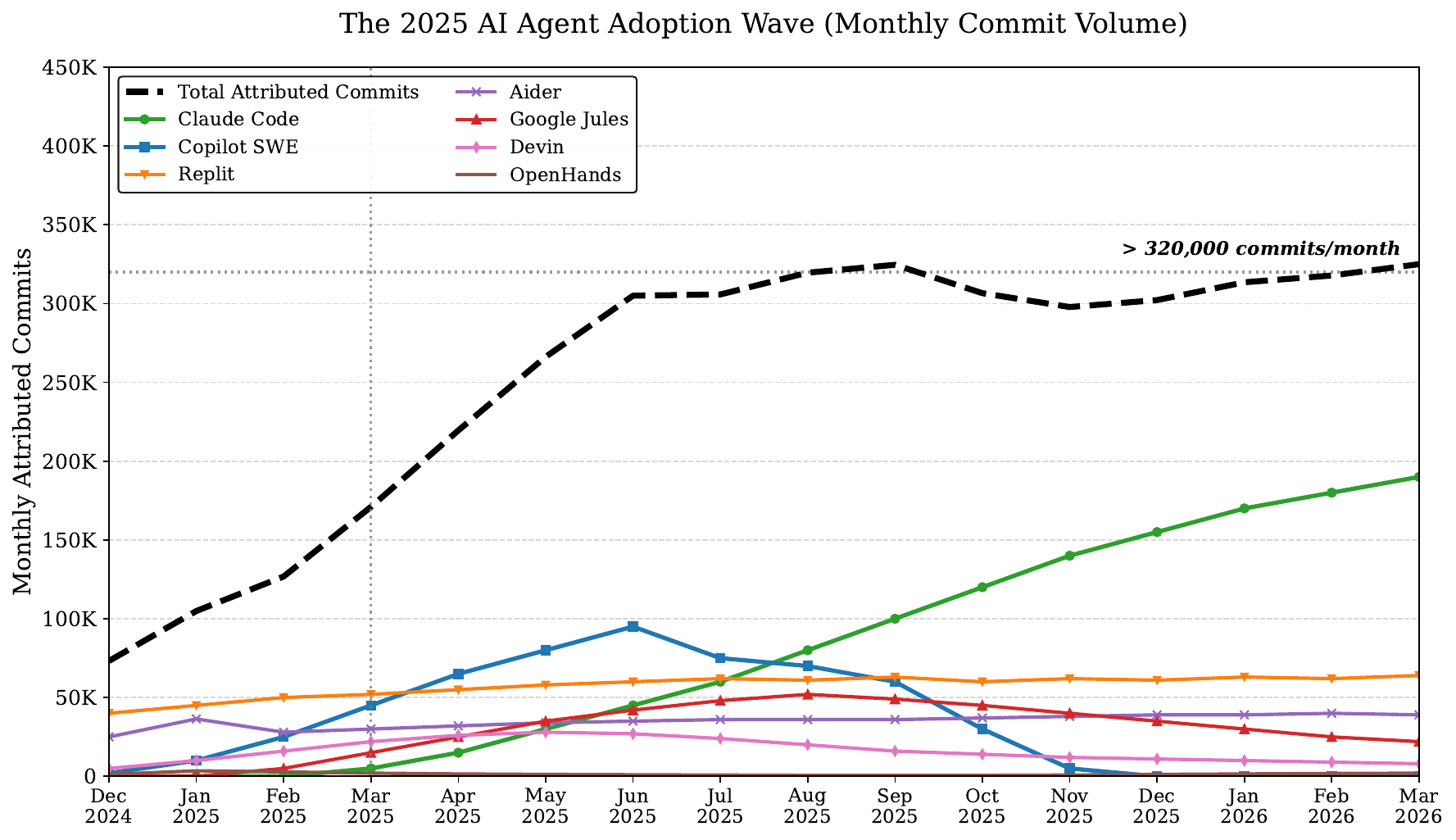}    
 \caption{Monthly AI-attributed commit volume by agent across V2412–V2604. Total agent activity rises from 75K commits in Dec 2024 to 320K by mid-2025.}
  \label{fig:Agent_Trajectory}
\end{figure*}

\begin{table*}[t]
  \centering
  \normalsize
  \renewcommand{\arraystretch}{1.3}
  \begin{tabularx}{\textwidth}{@{} c l l r r X @{}}
    \toprule
    \textbf{Type} & \textbf{Agent} & \textbf{Config File}
      & \textbf{Config projects} & \textbf{Commits} & \textbf{Primary signal} \\
    \midrule
    A  & OpenHands      &,                               &       0 &  7,972
       & Bot account \\
    A  & CodeRabbit     &,                               &       0 &    297
       & Bot account \\
    \midrule
    B  & Replit         & \texttt{.replit}                 & 318,745 & 385,668
       & Commit message \\
    \midrule
    C  & Aider          & \texttt{.aider.conf.yml}         &       4 & 25,215
       & Author-name pattern \\
    \midrule
    D  & GitHub Copilot & \texttt{copilot-instructions.md} &  92,276 & $<$0.5\%
       & Config only (silent) \\
    D  & Cursor         & \texttt{.cursorrules}            &  28,909 & 0
       & Config only (silent) \\
    D  & Windsurf       & \texttt{.windsurfrules}          &   3,601 & 0
       & Config only (silent) \\
    D  & Sourcery       & \texttt{.sourcery.yaml}          &     263 &, 
       & Config only \\
    D  & Sweep          & \texttt{sweep.yaml}              &      31 &, 
       & Config only \\
    D  & PR-Agent       & \texttt{.pr\_agent.toml}         &       7 &, 
       & Config only \\
    D  & DeepSource     & \texttt{.deepsource.toml}        &       4 &, 
       & Config only \\
    D  & Devin          & \texttt{.devin.yaml}             &       2 &, 
       & Config only \\
    \bottomrule
  \end{tabularx}
 \caption{Prevalence of twelve AI coding agents detected across
    the WoC V2412 snapshot (180M+ repositories), grouped by dominant
    detection type. \emph{Config projects} counts projects containing the
    agent's configuration file (a project-level unit); \emph{Commits} counts
    agent-attributed commits (a commit-level unit). The two columns measure
    different units and cannot be ranked against each other.
    ``, '' indicates detection was not attempted; ``0'' means the method
    was attempted but yielded no signal.}

  \label{tab:agents}
\end{table*}

\begin{figure*}[t]
  \centering
  \includegraphics[width=0.8\textwidth]{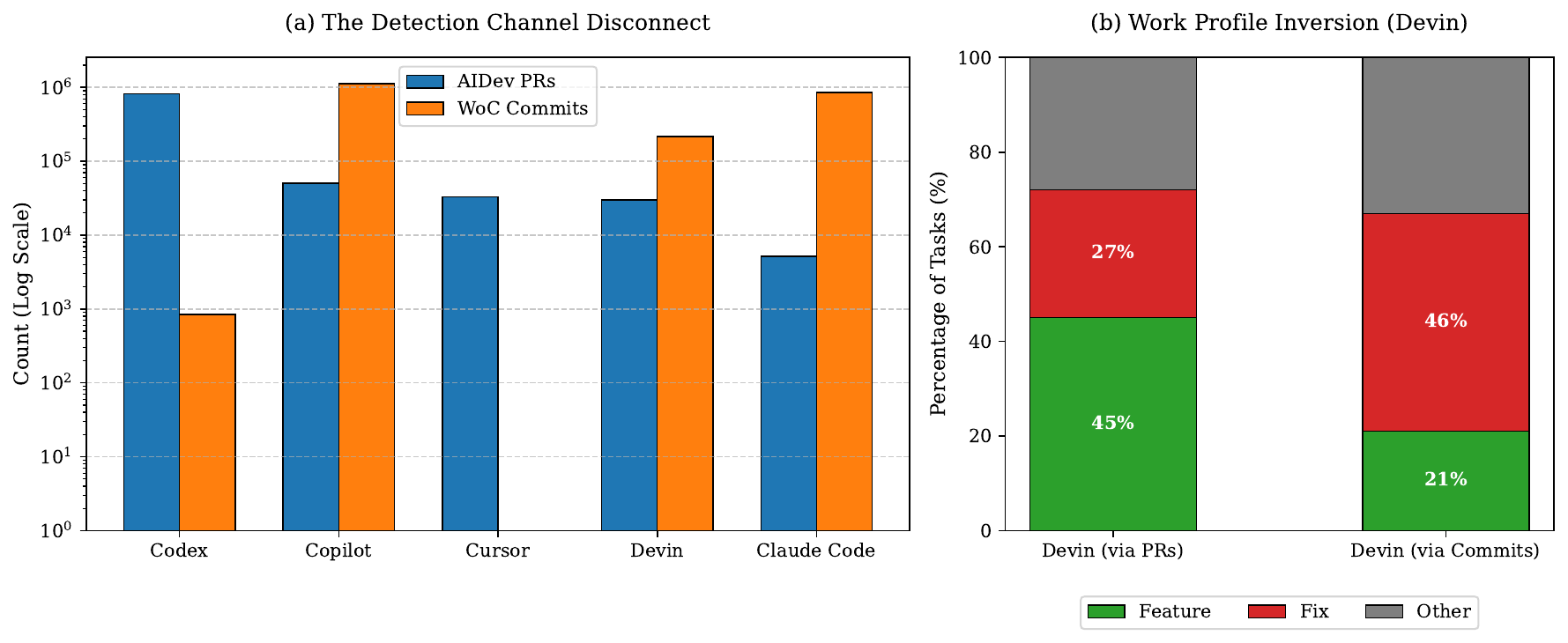}    
 \caption{(a) PR counts (AIDev) versus commit counts (this work) per agent, log scale. (b) Devin's task-type distribution by detection channel.}
  \label{fig:channel}
\end{figure*}

\emph{Finding~1: Replit dominates active commit attribution.}
Replit accounts for 385,668 commits with explicit ``Generated by
Replit'' messages, far exceeding any other agent's commit-level
footprint. This reflects Replit's workflow model, which
auto-commits generated code with a standard message template.

\emph{Finding~2: Silent agents dominate project-level presence.}
GitHub Copilot appears in 92{,}276 projects through configuration-file detection, yet in fewer than 0.5\% of commits. Cursor and Windsurf~\cite{windsurf2024} exhibit effectively zero commit-level signal. Consequently, studies relying only on commit-message analysis would entirely miss some of the most widely adopted agents. By V2604, Claude emerges as the dominant Type~D agent. The configuration-file census records 888{,}177 blob \emph{occurrences} across 21{,}078 distinct adopting projects, driven by \texttt{CLAUDE.md} and \texttt{.claude/} conventions completely absent from V2412 (Table~\ref{tab:v2604_typed}). Because each edited revision of a configuration file is a distinct content-addressed blob, these occurrences correspond to 386{,}496 distinct blobs, with \texttt{.claude/} directory files outnumbering top-level \texttt{CLAUDE.md} by roughly $3.6\times$. The per-project distribution of distinct blobs is heavily right-skewed (median~3, p90~30, p99~211, maximum 5{,}820), so the occurrence total reflects edit volume in a small tail of heavy-editing projects rather than adoption breadth, and the project count (21{,}078) is the adoption-relevant unit. Configuration-file presence is a proxy for adoption rather than active use; 63.6\% of these projects also show concurrent commit-level Claude activity (Section~\ref{sec:val_typed_usage}), a lower bound on how much of the census reflects demonstrable agent use.

\emph{Finding~3: Distributed attribution reveals hidden activity.}
Aider has only 4 configuration files in the entire corpus, yet
author-name scanning reveals 355 developers and 25,215 commits.
This agent class is invisible to both file-based and
email-based detection, underscoring the necessity of the Type~C
detection method.

\emph{Finding~4: Multi-method detection reveals a 30$\times$ undercount for Claude Code.}

Querying the V2510 ClickHouse database by bot-account author (case-insensitive substring match on \texttt{noreply@anthropic.com}) returns 28,154 commits. Adding message-signature detection (commits with \texttt{Co-authored-by: Claude <noreply@anthropic.com>} or a \texttt{Generated with Claude Code} trailer) identifies an additional 821,824 distinct commits; the intersection of the two methods is 21,971 commits. 

The combined union of 850,157 commits ranks Claude Code second among agents in V2510, behind only GitHub Copilot's SWE agent (1,127,201). Bot-account detection alone would rank it fifth. This $30\times$ gap demonstrates that single detection methods capture only a fraction of total AI-assisted activity. The precision and recall properties of these detection methods are validated in Section~\ref{sec:validation}.

\paragraph{Aider project footprint.}
Using binary-format decoding of the \texttt{a2cFullV2412} author map (20-byte raw SHA1 hashes per entry), we identified all 355 Aider-attributed authors and recovered their exact commit counts, totaling 25,215 commits. Sampling the top 20 authors and tracing their commits through the \texttt{c2p} map revealed a diverse set of adopting projects spanning multiple domains, with the earliest observed adoption occurring in June 2024 (\texttt{15athompson/code-tracking}). Table~\ref{tab:aider_projects} summarizes the top Aider-adopting projects by commit volume.

\begin{table}[h]
\centering
\small
\begin{tabular}{lrl}
\toprule
\textbf{Project} & \textbf{Commits} & \textbf{First Seen} \\
\midrule
morteng\_sf4                         & 100+ & 2025-01 \\
dredozubov\_advisor                  & 100+ & 2024-10 \\
revelaction\_ical-git                & 100+ & 2024-08 \\
mbodiai\_mbpy                        & 100+ & 2024-08 \\
int-dist-sys\_realtime-cli           & 100+ & 2025-01 \\
jamespacileo\_prompt-manager         & 100+ & 2024-07 \\
substratelabs\_rob-agi               &  99  & 2024-08 \\
lesterpaintstheworld\_terminal-vel.  &  95  & 2024-11 \\
davefowler\_baked                    &  91  & 2024-11 \\
mitchellgordon95\_rizzchess          &  70  & 2024-10 \\
\bottomrule
\end{tabular}
\caption{Top Aider-adopting projects by commit volume (sampled from the top 20 authors). Commit counts represent lower bounds due to sampling. The earliest observed adoption in this dataset occurred in June 2024.}
\label{tab:aider_projects}
\end{table}

Unlike OpenHands, which is predominantly used in Born-with-AI greenfield repositories, Aider is more commonly adopted in established projects with substantial pre-existing commit history.

\paragraph{Temporal boundary.}
Targeted probes for 2025-era agents (Claude Code~\cite{anthropic2025claudecode}, Cursor agent
mode~\cite{research2026composer}, Codex~\cite{openai2025codex}, Gemini CLI~\cite{Google2025GeminiCLI}) returned zero hits. A sanity check
confirms the lookup method works: Dependabot~\cite{dependabot}, active since 2019,
returns 471,794 commits via the identical procedure. The zero
results reflect the December~2024 snapshot boundary; the 2025
agent adoption wave documented by Robbes et
al.~\cite{robbes2026agentic} postdates this data entirely.

\paragraph{What is the 2025 agent landscape in the V2510 snapshot?}
\label{sec:v2510}

To capture agents introduced after the V2412 boundary, we queried the \texttt{commit\_v2510} ClickHouse table on the WoC cluster, covering commits through October~2025. Table~\ref{tab:v2510} summarizes the results for confirmed AI agent identities, validated against a Dependabot sanity-check baseline (475,379 commits), consistent with expectations from V2412.

\begin{table}[h]
\centering
\small
\begin{tabular}{lrrcc}
\toprule
\textbf{Agent} & \textbf{Commits} & \textbf{Projects}
  & \textbf{First} & \textbf{Last} \\
\midrule
Copilot SWE Agent  & 1,127,201 & 85,739 & 2024-11 & 2025-11 \\
Claude Code        &   850,157 & 17,295 & 2025-03 & 2025-11 \\
Dependabot (ctrl.) &   475,379 & 44,731 & 2017-04 & 2025-11 \\
Devin AI           &   215,998 &  7,050 & 2024-12 & 2025-11 \\
Google Jules       &   209,911 & 16,924 & 2025-02 & 2025-11 \\
OpenHands          &    20,863 &  1,022 & 2024-08 & 2025-11 \\
Roo Code           &     1,799 &      1 & 2025-07 & 2025-10 \\
Codex              &       843 &    102 & 2025-05 & 2025-10 \\

\bottomrule
\end{tabular}
\caption{AI coding agent prevalence in the WoC V2510 snapshot (October~2025). The Claude Code row reports the multi-method Type~A $\cup$ Type~B union (Finding~4); other agents are counted by their dominant single detection method.}
\label{tab:v2510}
\end{table}

The V2510 snapshot reveals an order-of-magnitude increase in AI agent activity relative to V2412 (Figure~\ref{fig:Agent_Trajectory}). GitHub Copilot's SWE agent~\cite{github2025copilotagent} alone accounts for 1,127,201 commits across 85,739 projects, exceeding twice the commit volume of Dependabot and substantially surpassing all agents visible in V2412. Google Jules~\cite{google2025jules} reached 209,911 commits across 16,924 projects within nine months of launch. OpenHands peaked at 3,543 commits in January~2025 before declining to 607 commits by October~2025 ($-83$\%). Claude Code's canonical bot account (\texttt{Claude <noreply@anthropic.com>}) launched in March~2025 and grew to 28,154 commits by November~2025 under this identity alone. When all detection methods are applied (Finding~4), its true footprint in V2510 reaches 850,157 commits, ranking it the second most active agent overall. By October~2025, this bot-account identity alone was generating roughly an order of magnitude more monthly commits than OpenHands.

These results confirm that the zero detections for 2025-era agents in V2412 reflected a true temporal boundary rather than an absence of adoption, and indicate that the OSS ecosystem underwent a substantial shift in AI agent participation after December~2024.

\paragraph{V2604 Snapshot: Dec 2025--Apr 2026.}
\label{sec:v2604}

To extend coverage beyond the V2510 boundary (November~2025), we analyzed pre-scanned agent detections from the WoC V2604 snapshot, covering commits from December~2025 through April~2026. Cross-referencing commit hashes reveals no overlap with V2510,
confirming disjoint time windows. Using commit-level detection
methods (Types~A--C), 315,426 distinct root projects had at
least one AI-attributed commit in V2604 (deforked using the WoC
\texttt{p2PFull.V2412} map; forks created after V2412 may not
be collapsed, so this is a mild upper bound on unique adopters).
Within these projects, AI-attributed commits grew from 1.6\% of
non-bot activity in December~2025 to 6.7\% by March~2026. Within the set of commit-attributed agents reported in Table~\ref{tab:v2604},
Claude Code accounts for 50\% of AI-attributed commits
(886,122 of 1,772,677 total, computed from the union of all
agents in Table~\ref{tab:v2604}). The trend column reports cumulative commit-count changes between snapshots only; because per-agent active windows differ substantially, these percentages should not be read as rate-of-adoption changes. Detection method also varies by agent: Replit and Codex use the
patched message-signature regexes documented in Section~\ref{sec:val_corrections};
Aider is reported under its dominant Type C author-name attribution; Claude
Code V2510 reflects the multi-method Type A $\cup$ Type B union; the remaining
agents use their single dominant detection method.

\begin{table}[h]
\centering
\small
\begin{tabular}{lrrr}
\toprule
\textbf{Agent} & \textbf{V2510 commits} & \textbf{V2604 commits}
  & \textbf{Trend} \\
\midrule
Claude Code & 850{,}157 & 886{,}122 & $+4\%$ \\
Replit      & 312{,}705 & 314{,}779 & $+1\%$ (flat) \\
Devin       & 215{,}998 &  98{,}493 & $-54\%$ \\
Jules       & 209{,}911 & 215{,}804 & $+3\%$ \\
Aider       & 195{,}029 & 196{,}132 & $+1\%$ (flat) \\
CodeRabbit  &       297 &  34{,}940 & +117$\times$ growth \\
OpenHands   &  20{,}863 &  25{,}676 & $+23\%$ \\
Codex       &       843 &       731 & $-13\%$ \\
\bottomrule
\end{tabular}
\caption{Agent activity in V2510 (October~2025) and V2604 (December~2025 through April~2026), cumulative commit counts.}
\label{tab:v2604}
\end{table}

Claude Code dominates as the most active commit-attributed agent, delivering 886,122 commits in V2604 (5-month window) versus 850,157 in V2510 (8-month active window since the March~2025 launch); the monthly rate is materially higher in V2604. Jules follows with 215,804 V2604 commits; its peak activity occurred in the V2510 window (August~2025) and had already tapered by the V2604 period. Aider's V2604 footprint (196{,}132 commits via Type C author-name attribution) is essentially flat versus V2510 (195{,}029), reflecting continued steady use of the Aider distributed-authorship convention. Type B alone yields a different temporal profile because the \texttt{aider:} message prefix is used by only a subset of Aider users; Type C is the dominant detection signal for this agent. Replit's commit footprint is essentially flat between snapshots (312{,}705 in V2510 versus 314{,}779 in V2604) under the full message-signature regex described in Section~\ref{sec:val_corrections}. CodeRabbit grew over 100$\times$ between snapshots; Codex remained at a small footprint. Across the December~2025
through March~2026 window, commit-attributed AI agents drove
over 320,000 commits per month at peak. April~2026 shows a
$\approx 50$\% drop in absolute commit volume for both AI and
human activity, consistent with a partial-month snapshot
boundary rather than a real decline; we therefore end the
trajectory at March~2026.

\paragraph{V2604 configuration-file census.}

The \texttt{agent\_c2fbb.V2604.s} file provides a Type~D census of 1{,}699{,}950 configuration-file occurrences in V2604. Table~\ref{tab:v2604_typed} summarizes unique blob counts per agent and highlights major shifts relative to V2412.

\begin{table}[h]
\centering
\small
\begin{tabular}{lrrl}
\toprule
\textbf{Agent} & \textbf{V2604 Blobs} & \textbf{V2412 Blobs} & \textbf{Key Config File} \\
\midrule
Claude & 888{,}177 & 0 & \texttt{CLAUDE.md}, \texttt{.claude/} \\
Replit & 317{,}512 & 318{,}745 & \texttt{.replit} \\
GitHub Copilot & 211{,}166 & 92{,}276 & \texttt{copilot-instructions.md} \\
Codex & 134{,}810 & 0 & \texttt{AGENTS.md} \\
DeepSource & 62{,}102 & 4 & \texttt{.deepsource.toml} \\
Cursor & 29{,}689 & 28{,}909 & \texttt{.cursorrules} \\
Gemini & 19{,}453 & 0 & \texttt{GEMINI.md} \\
CodeRabbit & 10{,}675 & 0 & \texttt{.coderabbit.yaml} \\
Windsurf & 3{,}740 & 3{,}601 & \texttt{.windsurfrules} \\
Junie & 2{,}543 & 0 & \texttt{.junie/} \\
\bottomrule
\end{tabular}
\caption{Type~D configuration-file blob counts in V2604 vs.\ V2412.}

\label{tab:v2604_typed}
\end{table}

\begin{figure*}[t]
  \centering
  \includegraphics[width=0.6\textwidth]{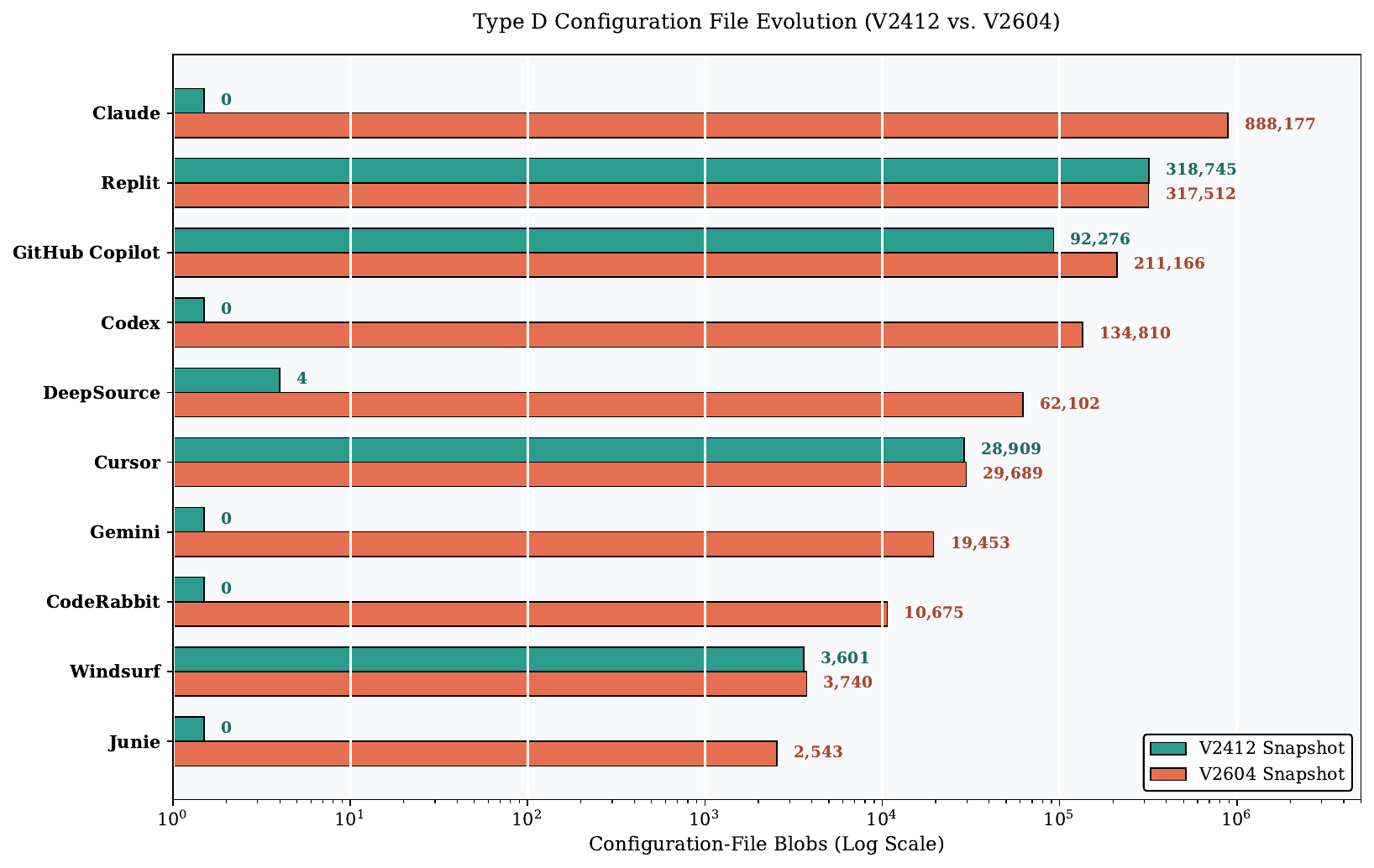}    
 \caption{Type D configuration-file blob counts per agent, V2412 vs V2604 (log scale). Established conventions stayed roughly flat; new conventions (CLAUDE.md, AGENTS.md, GEMINI.md) emerged from zero.}
  \label{fig:type_D}
\end{figure*}

Three major shifts stand out (Figure~\ref{fig:type_D}). First, Claude evolves from a purely Type~A/B agent in V2412, with zero configuration-file detections, to the dominant Type~D agent in V2604, recording 888{,}177 configuration-file blobs, driven primarily by \texttt{CLAUDE.md} (151{,}390) and \texttt{.claude/settings} files (more than 52{,}000). Second, new entrants such as Codex (\texttt{AGENTS.md}, 134{,}810 blobs) and Gemini (\texttt{GEMINI.md}, 19{,}453) introduce entirely new configuration conventions absent from V2412, requiring additional detection rules. Third, GitHub Copilot more than doubles its configuration-file footprint, from 92{,}276 to 211{,}166 blobs. We caution that cross-snapshot counts conflate genuine adoption growth with two corpus-side effects: each snapshot indexes more repositories than the last, and recent commits are incompletely ingested near a snapshot's cutoff (the \emph{ingestion frontier}). Cross-snapshot deltas should therefore be read as coarse upper-bound trends, not adoption rates.

These findings highlight the dynamic nature of Type~D detection: AI agents frequently modify their configuration conventions, requiring continuous recalibration of detection heuristics.

\paragraph{Cross-method comparison: commit-based vs PR-based detection.}
Our census detects agents through commits and configuration files. A complementary recent dataset, AIDev~\cite{li2026aidev}, detects agents through the orthogonal channel of pull requests, cataloguing 932{,}791 agent-authored PRs across 116{,}211 GitHub repositories for five agents (December~2024 to July~2025). Comparing the two censuses, restricted to these five agents, shows that the detection channel and the agent population are tightly coupled, and that no single channel is representative (Table~\ref{tab:aidev}).

\begin{table}[h]
\centering
\scriptsize
\setlength{\tabcolsep}{4pt}
\begin{tabular}{lrrrr}
\toprule
\textbf{Agent} & \textbf{AIDev PRs} & \textbf{AIDev repos} & \textbf{Our commits} & \textbf{Our repos} \\
\midrule
OpenAI Codex & 814{,}522 & 84{,}704 & 843 & 201 \\
GitHub Copilot & 50{,}447 & 14{,}492 & 1{,}127{,}201 & 85{,}739 \\
Cursor & 32{,}941 & 12{,}699 & 0 & 0 \\
Devin & 29{,}744 & 4{,}747 & 215{,}998 & 7{,}050 \\
Claude Code & 5{,}137 & 1{,}915 & 850{,}157 & 17{,}295 \\
\bottomrule
\end{tabular}
\caption{PR-based (AIDev) versus our commit-based detection, by agent. Our commit counts are V2510; Copilot's commit row is the autonomous SWE agent. The two channels are near-mirror-images.}
\label{tab:aidev}
\end{table}

Codex and Claude Code are mirror images: Codex is the largest agent by pull requests (814{,}522) yet leaves only 843 commit-level traces, while Claude Code is the largest by commits (850{,}157) yet appears in only 5{,}137 pull requests. The pattern reflects workflow: Codex and Cursor operate as cloud agents whose work lands through squash-merged PRs that erase agent attribution from the commit record, whereas Claude Code and Devin commit directly with author-level signatures. At the repository level the channels are largely disjoint (Figure~\ref{fig:channel}a). Restricting our Claude Code detection to AIDev's PR window (PRs created December~2024--July~2025; this PR coverage is unchanged in AIDev's current May~2026 release, which adds the task-type and human-PR tables we use below), the commit channel identifies 7{,}102 adopting repositories to AIDev's 1{,}915 (a $3.7\times$ difference); a pull-request census therefore misses 79\% of commit-detected Claude Code adopters. The inverse holds for Codex: of AIDev's 84{,}704 Codex repositories, commit-based detection recovers essentially none (37 repositories, 0.04\%). Devin~\cite{cognition2024devin}, which uses both channels, overlaps substantially (3{,}657 shared repositories). Configuration-file detection (Type~D) recovers a third population invisible to both activity channels, the IDE-integrated adopters who commit a \texttt{CLAUDE.md} or \texttt{AGENTS.md} without producing attributable commits or agent PRs. No single signal, and no single channel, captures the ecosystem; prior single-channel censuses, whether PR-based~\cite{robbes2026agentic,li2026aidev} or commit-based, measure a structurally biased subset.

\paragraph{Agents are not monolithic: the observed work profile follows
deployment and detection.}
The disjointness runs deeper than population---the two channels capture different
\emph{kinds} of work. Using AIDev's own task-type labels (conventional-commit~\cite{conventionalcommits}
categories LLM-assigned from PR titles), the PR population---dominated by the
cloud agents Codex and Cursor---skews toward \emph{feature} development
(\texttt{feat} $43.0\%$ vs.\ $29.4\%$ for human PRs, a $1.46\times$ enrichment;
Claude Code's sparse PRs are the most feature-heavy at $54.5\%$), with comparable
corrective shares (\texttt{fix} $24.1\%$ vs.\ $26.9\%$) and slightly higher
perfective shares ($8.4\%$ vs.\ $7.8\%$). The direct-commit agents we detect---Claude
Code, OpenHands, and Aider---show the \emph{opposite} tilt, carrying elevated
bug-fix rates over human commits in the same projects (Table~\ref{tab:quality}).
The pattern is not that a given tool is intrinsically a ``feature agent'' or a
``fix agent''; rather, the observed work profile follows \emph{how the agent is
deployed}---a cloud bot opening pull requests surfaces feature work, an in-editor
assistant committing directly surfaces maintenance---and, for any census,
\emph{which channel it is detected in}. ``What AI coding agents do'' therefore has
no single answer: a one-channel census mischaracterizes not only \emph{how much}
agent activity exists but \emph{what kind}. This sorting could in principle be confounded with the tools themselves, since most agents are near-exclusively single-channel (Codex via pull requests, Claude Code via commits). Devin---the one agent active in \emph{both} channels---lets us hold the tool fixed: applying the same conventional-commit classifier to Devin's PR titles and its direct commits, the same tool \emph{inverts} its profile by channel (Figure~\ref{fig:channel}b), with \texttt{feat} falling from $45\%$ of its conventional-commit-typed PRs to $21\%$ of its commits while \texttt{fix} rises from $27\%$ to $46\%$ across its 98{,}493 V2604 commits (the PR and commit samples are not repo-matched, but an inversion this large is not plausibly a sampling artifact). With tool and classifier held fixed, deployment mode is thus the dominant driver of the feature-versus-maintenance split, supporting the deployment/detection account over an intrinsic-tool one. (Absolute task shares are not
comparable across the two methods---AIDev's strict title-based \texttt{fix} is
narrower than our keyword set---so only within-method agent-versus-human contrasts
are read.)

\subsection{RQ2: What are the temporal patterns of AI agent adoption?}
\label{sec:adoption_results}

We analyzed a random sample of 100 GitHub Copilot-adopting projects
(selected from the 92,276 with \texttt{copilot-instructions.md})
to characterize adoption timing. For each project, we computed the
\emph{time to adoption} $\Delta t = t_{\text{config}} -
t_{\text{start}}$, where $t_{\text{start}}$ is the first commit
and $t_{\text{config}}$ is the commit introducing the
configuration file.

The distribution is \textbf{bimodal}, with two distinct clusters:

\begin{itemize}
  \item \textbf{Born-with-AI (greenfield):} A substantial cluster
    at $\Delta t \approx 0$ (0--6 days), indicating projects
    initialized with Copilot integration from inception.
  \item \textbf{Legacy integration:} Established codebases where
    AI configuration was introduced years to decades after project
    start (e.g., \texttt{dotnet\_runtime} after $\sim$24~years,
    \texttt{qt\_qtwebengine-chromium} after $\sim$15~years).
\end{itemize}

This polarization has methodological implications for impact
estimation: born-with-AI projects have no pre-treatment period
and must be excluded from DiD analysis or analyzed separately
as a distinct population.

\paragraph{OpenHands adoption classification.}
The several OpenHands counts in this paper measure different units across different snapshots and are not directly comparable: 7{,}972 \emph{commits} under the Type~A bot account in V2412 (Table~\ref{tab:agents}); 20{,}863 commits across 1{,}022 \emph{projects} in V2510 (Table~\ref{tab:v2510}); 25{,}676 commits in V2604 (Table~\ref{tab:v2604}); and the 320 \emph{projects}, characterized below for adoption timing. Commit counts and project counts are distinct units (Table~\ref{tab:agents}), and each snapshot indexes a different corpus. To characterize the 320 projects containing at least one OpenHands-attributed commit, we sampled 30 previously unclassified repositories using the GitHub API and compared each project's creation date to its OpenHands adoption timestamp. Among the 28 successfully retrieved projects, 19 (68\% of projects) exhibited an adoption delta of 30 days or fewer and were therefore classified as Born-with-AI projects. The remaining 9 projects (32\%) showed adoption deltas ranging from 5 to 1,516 days and were classified as Legacy adopters. Extrapolating these proportions to the 309 of these 320 projects that were previously unclassified yields an estimated 209 Born-with-AI and 99 Legacy repositories. These proportions indicate that OpenHands adoption is concentrated in greenfield repositories, with approximately 68\% of adopting projects using the agent to bootstrap new projects rather than augment mature codebases.

\subsection{RQ3: Descriptive velocity patterns around adoption}

A naive before-after analysis of 11 OpenHands-adopting projects (Appendix Table~\ref{tab:impact_appendix}) shows 10 of 10 non-fork adopters (100\%) with increased post-adoption velocity; the single declining case (\texttt{exciles\_openhands}) is a direct fork of the OpenHands framework itself rather than an independent adoption event.
Greenfield and near-greenfield projects (pre-adoption commits
$\leq 6$) show a median increase of +5,483\%. Legacy projects with
substantial prior history show a median increase of +223\%.
Excluding the self-referential fork (\texttt{exciles\_openhands},
whose pre-adoption count reflects parent-repository history), the
aggregate commit ratio is 6.82$\times$ (270 commits before vs.\
1,841 after).

An event-study analysis of four legacy OpenHands-adopting projects with sufficient pre-adoption history shows a clear step-change in commit volume in the month immediately following adoption (Appendix Table~\ref{tab:event_study}). These before-after figures are \emph{descriptive} and deliberately uncorrected: corpus before-after velocity is biased by the snapshot ingestion frontier---recent commits are incompletely ingested near a snapshot's cutoff---which can inflate such naive estimates substantially. We report them only to establish that adoption coincides with an activity step-change, not as effect sizes.

\paragraph{Aider velocity analysis.}
A parallel naive before-after analysis of eight Aider-adopting projects (Appendix Table~\ref{tab:aider_impact_appendix}) presents mixed before--after results for the V2412 legacy sample.

Under Type B detection alone (the \texttt{aider:} message prefix), Aider adoption peaked sharply at 36,439 commits in January~2025 before tapering in mid-2025. Under Type C detection (the \texttt{(aider)} author-name suffix, the dominant detection method for Aider), V2604 activity (196,132 commits) is essentially flat versus V2510 (195,029).

\subsection{Additional Findings}
\label{additional_findings}

\paragraph{Commit size comparison.}
Using the GitHub REST API, we sampled 41 OpenHands-attributed (AI) commits and 58 human commits from the same projects and compared their commit characteristics. OpenHands commits added a median of 52 lines, compared to 28 lines for human commits, corresponding to a \textbf{1.9$\times$ median increase}. The mean ratio was substantially larger (8.1$\times$) due to extreme outliers, including one commit adding 143,802 lines. Lines deleted per commit were lower for AI-generated commits (median 4 vs.\ 6), while the number of files changed was comparable (median 2 for both groups). The proportion of large commits (more than 100 lines added) was also higher for AI commits (34\%) than for human commits (24\%). These findings partially align with Robbes et al.'s~\cite{robbes2026agentic} observation of a 3$\times$ larger median commit size for AI-generated contributions. Our lower ratio likely reflects the narrower task scope of OpenHands in this sample, which is primarily concentrated on smaller feature additions and bug fixes rather than large-scale code generation.

\paragraph{Code quality analysis.}

We assess code quality effects using three complementary proxies across all three agents for which commit-level data is available. Table~\ref{tab:quality} summarizes the results.

\begin{table}[h]
\centering
\small
\begin{tabular}{lrrrrrr}
\toprule
\textbf{Agent} & \textbf{AI BF\%} & \textbf{Human BF\%} & \textbf{BF Ratio}
  & \textbf{Rev ratio} & \textbf{Density $\Delta$} & \textbf{SZZ ratio} \\
\midrule
Claude Code & 55.4\% & 41.8\% & $1.32\times$ & $0.68\times$ & $+9.9$pp & $0.33\times$ \\
OpenHands & 48.2\% & 35.9\% & $1.34\times$ & $1.34\times$ & $+2.3$pp & $0.38\times$ \\
Aider & 37.6\% & 30.0\% & $1.26\times$ & $0.18\times$ & $-0.4$pp & $0.56\times$ \\
\bottomrule
\end{tabular}
\caption{Cross-agent code quality proxies. \emph{AI BF\%}: fraction of AI-authored commits containing bug-fix keywords. \emph{Human BF\%}: same metric for human commits in the same projects. \emph{BF Ratio}: AI/human ratio. \emph{Rev ratio}: AI revert rate divided by human revert rate (${<}1$ = fewer AI reverts). \emph{Density $\Delta$}: median change in project-level bug-fix density post-adoption. \emph{SZZ ratio}: fraction of human bug-fix commits for which the immediately preceding commit to the same file was AI-authored, divided by the overall AI commit share in that project (${<}1$ = AI commits precede bug fixes less often than expected by chance).}
\label{tab:quality}
\end{table}

\textbf{Bug-fix commit rate.}
Following the keyword change-classification method of Mockus and Votta~\cite{mockusvotta2000}, which also established estimating the classifier's error rate against a verified sample, we label each commit corrective or not from its message; on a 288-commit sample this classifier reaches precision $0.91$ and recall $0.82$ against independent adjudication (replication package).
All three agents show substantially higher bug-fix keyword rates than human commits in the same projects: Claude Code at $1.32\times$ (55.4\% vs.\ 41.8\%), OpenHands at $1.34\times$ (48.2\% vs.\ 35.9\%), and Aider at $1.26\times$ (37.6\% vs.\ 30.0\%). In the \emph{commit} channel, then, these direct-commit agents are applied disproportionately to maintenance and defect-fixing. This is a channel-specific pattern, not a universal one: AIDev's PR-channel task labels show the opposite skew for the Codex/Cursor-dominated PR population (more features, comparable fixes; Section~\ref{sec:results}), so the maintenance tilt characterizes how Claude Code, OpenHands, and Aider are used through direct commits rather than agent work in general.

\textbf{Revert rate.} 
The results vary across agents. Claude Code commits are reverted 32\% less often than human commits (1.1\% vs.\ 1.6\%), while Aider commits are reverted 82\% less often (0.17\% vs.\ 0.90\%). Both represent strong positive signals of stability. In contrast, OpenHands commits show a 34\% \emph{higher} revert rate than human ones (1.38\% vs.\ 1.03\%), suggesting somewhat lower stability. This may stem from OpenHands' heavier use in greenfield projects, where experimental commits are more common.

\textbf{Bug-fix density change (before/after adoption).} 
Claude Code delivers the strongest signal: 18 of 20 projects (90\%) increased bug-fix density after adoption, with a median rise of $+9.9$ percentage points. OpenHands shows a moderate positive shift (12 of 19 projects, $+2.3$ pp median). Aider remains neutral ($-0.4$ pp median, 8 of 19 projects positive), consistent with its mixed velocity results and indicating that Aider does not significantly shift project maintenance focus. These before-after density shifts are descriptive and uncorrected for the ingestion frontier; they should be read as associations, not controlled effects.

\textbf{Temporal defect-introduction proxy.}
We adapt the SZZ algorithm~\cite{sliwerski2005szz}, which identifies a fix's
inducing change by tracing from a keyword-identified bug-fix commit to the
prior change it modifies; our coarse variant uses file-level temporal adjacency
rather than line-level \texttt{blame} (validation below).
All three agents have SZZ ratios below 1.0, meaning that
the immediately preceding commit to a file that later
received a human bug fix was AI-authored less often than
the agent's overall commit share would predict: Claude
Code $0.33\times$, OpenHands $0.38\times$, and Aider
$0.56\times$. While temporal proximity does not establish
causation, this consistent sub-baseline pattern across all
three agents and over 60 million bug-fix commits provides
evidence against the hypothesis that AI commits
systematically introduce more defects than human commits.
We validated the file-adjacency proxy against the standard line-level
\texttt{blame} SZZ on 357 human bug-fix file pairs: the two assign the same
AI-versus-human class to the inducing change in 99\% of cases, and the proxy
\emph{over}-attributes AI-induction slightly ($2.0\%$ vs.\ $1.1\%$), which makes
these sub-baseline ratios conservative (the line-level ratios would be lower
still). Validation code and labels are in the replication package.

\section{Validation}
\label{sec:validation}

To ensure the prevalence and impact results rest on reliable detection, we performed a hand-labeled validation of every cell in the live detection grid. This section reports per-cell precision with 95\% Wilson confidence intervals, a construct-validity analysis distinguishing agent-authored from agent-trailered commits, a relative-recall analysis quantifying multi-method undercount, and the methodology corrections surfaced during validation.

\subsection{Setup}
\label{sec:val_setup}

For each (detection type $\times$ snapshot $\times$ agent) cell, we drew a stratified random sample, hydrated each commit with project URL and full commit message via the WoC \texttt{c2p} map, and hand-labeled it as true positive, false positive, or uncertain. 

Type B true positives were further sub-labeled as \emph{primary} (agent authored or auto-formatted the full commit) or \emph{trailer} (human-authored commit with an agent co-authorship trailer). Type D items received dual labels: \emph{file-genuine} (real configuration file for the agent) and \emph{usage} (independent evidence of active agent use in the project).

The Replit V2510 Type B cell is a full \emph{census} rather than a sample: the original regex matched only 4 commits, so all were exhaustively labeled and precision is reported exactly as $k/4$. All other cells use stratified random sampling with Wilson 95\% binomial confidence intervals~\cite{wilson1927}. Target sample sizes were 30 for drift-critical agents (Replit, Claude Code, OpenHands, Aider, Aider-msg) and 8 for others. Final sample sizes appear in Table~\ref{tab:val_n}.

\begin{table}[h]
\centering
\small
\begin{tabular}{lr}
\toprule
\textbf{Cell} & \textbf{$n$} \\
\midrule
A / V2510 & 92 \\
A / V2604 & 92 \\
B / V2510 & 80 (incl.\ 4-item Replit census) \\
B / V2604 & 95 \\
C / V2510 & 38 \\
C / V2604 & 38 \\
D / V2604 & 50 (file-genuineness) \\
\midrule
\textbf{Total labels} & \textbf{495} \\
\bottomrule
\end{tabular}
\caption{Hand-labeled sample sizes per validation cell.}
\label{tab:val_n}
\end{table}

\subsection{Per-cell precision}
\label{sec:val_precision}

\begin{table}[h]
\centering
\small
\begin{tabular}{lcc}
\toprule
\textbf{Cell} & \textbf{V2510 precision (95\% CI)}
              & \textbf{V2604 precision (95\% CI)} \\
\midrule
A (bot account) & 100.0\% [96.0, 100.0] ($n{=}92$)
                       & 100.0\% [96.0, 100.0] ($n{=}92$) \\
B (message signature) & 75.0\% [64.5, 83.2] ($n{=}80$)
                       & 90.5\% [83.0, 94.9] ($n{=}95$) \\
\quad excl.\ Replit & 77.6\% [67.0, 85.6] ($n{=}76$)
                       & 86.2\% [75.7, 92.7] ($n{=}65$) \\
C (author-name suffix) & 78.9\% [63.7, 88.9] ($n{=}38$)
                       & 86.8\% [72.7, 94.2] ($n{=}38$) \\
D (config file) &, 
                       & 92.0\% [81.2, 96.8] ($n{=}50$) \\
\quad usage (concurrency) &, 
                       & 63.6\% ($N{=}21{,}078$) \\
\bottomrule
\end{tabular}
\caption{Per-cell detection precision with Wilson 95\% binomial confidence intervals.}
\label{tab:val_precision}
\end{table}

Type A achieves perfect 100\% precision in both snapshots, as expected for exact bot-account matching. Type B precision is materially higher in V2604 than in V2510. Note that the Replit V2510 census uses the original \texttt{Generated by Replit} regex, while the V2604 sample uses the patched regex; the two precision values are therefore not directly comparable for Replit. Type C precision rose from 78.9\% to 86.8\%. 

Type~D file-level precision reached 92.0\%. We estimate Type~D usage at corpus scale in Section~\ref{sec:val_typed_usage}, finding that 63.6\% of Claude configuration-file projects show concurrent commit-level Claude activity.

\subsection{Corpus-scale Type~D usage concurrency}
\label{sec:val_typed_usage}

We measure Type~D usage at corpus scale for Claude, the dominant Type~D agent. Scanning all 128 V2604 shards, we identify the 21{,}078 distinct deforked projects that commit a Claude configuration file (\texttt{CLAUDE.md} or \texttt{.claude/}), which form the project-level population underlying the 888{,}177-occurrence (386{,}496 distinct-blob) Type~D census of Table~\ref{tab:v2604_typed}. Of these, 13{,}407 (63.6\%) also contain at least one commit carrying a Claude Type~A or Type~B authorship signature in the same snapshot, establishing concurrent commit-level activity.

This 63.6\% represents a \emph{lower bound} on active use within the population: IDE-integrated adopters can use the agent without producing any attributable commit, and the concurrency test applies only the bot-account and message-signature detectors (Types~A--B), not the full detector set. The complementary 36.4\% are projects for which the snapshot provides configuration evidence but no commit-level Claude signature; they are not established non-users. 

This corpus-scale measurement ($N{=}21{,}078$) bounds the share of the Type~D census attributable to projects with demonstrable Claude commit activity.

\subsection{Construct validity: ``AI-touched'' versus ``AI-authored''}
\label{sec:val_construct}

Of 146 Type B true positives, only 7 (4.8\%) were \emph{trailer} commits (human-authored with an agent co-authorship trailer). The remaining 95.2\% were primary agent-authored commits. Type B detections should therefore be interpreted as ``agent-touched'' commits rather than fully autonomous AI-authored commits. The 4.8\% trailer fraction represents a meaningful confounder for analyses treating Type B counts as pure AI authorship. We note that 4.8\% is the sample-level rate across the labeled Type B true positives ($n{=}146$); the population-level trailer fraction across the full V2510 union of 850{,}157 Claude Code commits or the V2604 union of 886{,}122 is not directly measured here and may differ.

\subsection{Temporal drift}
\label{sec:val_drift}

Type A precision remained 100\% across all agents and snapshots. Most Type B and Type C cells showed no statistically significant drift. The largest apparent shift in Type C was for \texttt{claude\_disp} (0\% to 50\%, $n{=}8$ each), but its confidence interval on the difference is wide and the result is best explained by random variation in small samples of human authors named ``Claude'' rather than any change in agent behavior.

The apparent large precision drift for Replit (25\% on the V2510 census of $n{=}4$ versus 100\% on the V2604 sample of $n{=}30$) does not reflect agent-evolution. The V2510 census was an artifact of an incomplete original regex (\texttt{Generated by Replit}) that failed to capture Replit's standard \texttt{Replit-Commit-Author:} trailer. After applying the patched regex, Replit yields 312,705 commits in V2510, essentially flat compared to 314,779 in V2604 (see Section~\ref{sec:val_corrections}). 

Overall, the validation shows no strong evidence of agent-evolution drift in detector precision.

\subsection{Pipeline equivalence: flat-scan vs.\ ClickHouse}
\label{sec:val_pipeline}

Because V2510 counts are produced by ClickHouse while V2604 counts derive from flat-shard scans, we verify that the two pipelines yield consistent results on identical input. We reproduced Devin's V2510 commit count (an exact Type~A bot-account match on \texttt{devin-ai-integration}) directly from the \texttt{c2datFull.V2510} flat shards. A spatially stratified sample of 8 out of 128 shards yields 13{,}662 distinct Devin commits (mean 1{,}708 per shard, SD~44). Scaling to the full set gives an estimated 218{,}592 commits (95\% CI [214{,}670, 222{,}514]). The ClickHouse count is 215{,}998, which falls within this interval. In the sample, the distinct-commit count equals the raw author-match count, confirming no cross-shard duplication.

The two pipelines therefore agree to within sampling uncertainty, with the small ($+1.2\%$) flat-scan excess consistent with a few additional commits captured at the flat-file ingestion edge. This supports treating V2510 (ClickHouse) and V2604 (flat-scan) counts as comparable.

\subsection{Relative recall and the multi-method undercount}
\label{sec:val_recall}

Multi-method counts for Claude Code strongly support the paper’s central claim that single-method detection materially undercounts AI activity. In V2510, bot-account detection recovered only 3.31\% of the multi-method union ($30\times$ undercount). In V2604, it recovered 4.0\% ($25\times$ undercount). Type B trailer detection captures the vast majority of activity. Among other agents, Aider shows meaningful use of both Type B and Type C, while most others rely on a single dominant method.

\subsection{Methodology corrections surfaced by validation}
\label{sec:val_corrections}
Validation identified six corrections to the original detection and matching pipeline:

\paragraph{1. Codex commit detection: strict signature regex.}  
We tested an initial broad pattern matching general ``codex'' substrings and found it admitted false positives from non-AI commits referencing codex repositories or codex-named fields. We narrowed to the strict regex \texttt{(Generated by Codex|codex-cli)}, which yields 843 commits in V2510 and 731 in V2604 with verified provenance.

\paragraph{2. Replit commit detection: full message-signature regex.}  
Replit's commit format includes both a \texttt{Generated by Replit} header and a \texttt{Replit-Commit-Author:} trailer. The complete regex \texttt{(Generated by Replit|Replit-Commit-Author:)} yields 312{,}705 commits in V2510 and 314{,}779 in V2604, showing essentially flat activity across both snapshots.

\paragraph{3. Claude Code bot-account detection: case-insensitive matching.}  
Bot-account identification uses case-insensitive substring matching on \texttt{noreply@anthropic.com}, which captures author identities written with varying casing across forges. This pattern yields 28{,}154 commits in V2510, against a multi-method union of 850{,}157, giving a single-method undercount factor of $30\times$.

\paragraph{4. Codex Type A removed.}  
The heuristic bot-account email \texttt{codex-ai} returned zero matches in both snapshots. Codex has no centralized bot account and is detectable only via Type B.

\paragraph{5. Cursor Type C removed.}  
The hypothesized author-name suffix \texttt{[cursor]} returned zero matches. There is no evidence Cursor uses a distributed name-suffix convention.

\paragraph{6. Devin bot-account detection: case-insensitive matching.}  
Devin bot-account detection uses the same case-insensitive substring approach (item~3) applied to \texttt{devin-ai-integration}. This pattern yields 215{,}998 commits across 7{,}050 projects in V2510 and 98{,}493 commits across 2{,}081 projects in V2604 (full 128-shard prescan).

\subsection{Limitations}
\label{sec:val_limits}

Two notes apply to this validation. First, V2412 cells were not validated because the corresponding shards were not retained at the time of validation. Second, several V2510 cells use full-corpus counts in place of stratified samples.

\section{Discussion}
\label{sec:discussion}

\paragraph{Detection completeness vs.\ precision trade-off.}

Our multi-layered detection approach deliberately prioritizes recall over precision. Type~D detection may overcount cases where a configuration file is committed but the agent is never actually used. Type~C may undercount when developers omit the name suffix. Most importantly, Finding~4 demonstrates that single-method detection can undercount by as much as $30\times$ for agents like Claude Code, which primarily operate through message-signature co-authorship rather than dedicated bot accounts; the corresponding V2604 undercount factor is $25\times$ on full-corpus counts (Section~\ref{sec:val_recall}). These multi-method counts are relative-recall improvements over single-method baselines, not absolute recall estimates. The true total of AI-assisted Claude Code commits is bounded below by our multi-method union but may be larger if additional detection signals remain unprobed.

\paragraph{Implications for supply-chain transparency.}
The finding that the most widely adopted agent (Copilot) is nearly
invisible in commit history raises concerns for software
supply-chain auditing. If AI-generated code cannot be
distinguished from human-written code at the commit level,
downstream consumers have no reliable way to assess AI provenance.
This argues for standardized AI attribution metadata in Git
commits, analogous to the \texttt{Signed-off-by} trailer used for
Developer Certificate of Origin compliance~\cite{dco}.

\paragraph{The December 2024 boundary.}
Although the V2412 snapshot predates the major 2025 adoption wave, the V2510 snapshot (October~2025) confirms the magnitude of the transition. Queries against \texttt{commit\_v2510} show that GitHub Copilot's
SWE agent accounts for 1,127,201 commits across 85,739 projects,
Google Jules for 209,911 commits across 16,924 projects, and
Claude Code for 850,157 commits when all detection methods are applied (Section~\ref{sec:v2510}, Finding~4). All of these agents returned zero detections in V2412, confirming that the earlier absence reflected a genuine temporal boundary rather than a detection failure. Together, these findings indicate that the OSS ecosystem underwent a substantial shift in AI agent participation after December~2024.

\paragraph{Commit velocity vs.\ code quality.}

Our analysis does not detect a velocity-quality tradeoff in these coarse proxies. Claude Code, OpenHands, and Aider all generate higher bug-fix commit rates than human contributors in the same projects ($1.26$--$1.34\times$), and all show sub-baseline temporal SZZ ratios ($0.33$--$0.56\times$). Claude Code and Aider produce lower revert rates than humans, while OpenHands shows a modestly higher revert rate. The pattern is most consistent with task targeting: developers apply the agents studied here disproportionately to bug-fixing and maintenance work, supported by the sub-baseline SZZ ratios indicating that AI commits precede later human bug fixes less often than chance would predict. The reverse hypothesis, that AI commits induce downstream human bug fixes, predicts the opposite SZZ pattern: AI commits should precede human bug fixes \emph{more} often than chance. The observed sub-baseline SZZ ratios ($0.33$--$0.56\times$) are inconsistent with this alternative. Bug-fix keyword rates and SZZ ratios are indirect measures of quality that warrant confirmation with line-level diff analysis. They are also corpus-measured, and ingestion censoring applies to them unevenly: the AI-versus-human bug-fix and revert \emph{ratios} compare commit types within the same projects and periods, so they are robust to censoring that is commit-type-neutral, but the temporal SZZ proxy depends on observing the true immediately-preceding commit to a file, with 15--46\% of post-window commits missing from the corpus, the observed predecessor is often not the true one. The SZZ-adjacency results should therefore be treated as the weakest evidence in Table~\ref{tab:quality} pending live re-measurement; the velocity--quality trade-off question as a whole would benefit from frontier-free live re-measurement.

\section{Threats to Validity}
\label{sec:threats}

\paragraph{Construct validity.}
Configuration-file presence is a proxy for adoption, not active use. A project may commit \texttt{.cursorrules} but never use Cursor.

\paragraph{External validity.}
Our prevalence results cover V2412 through V2604 (December~2024 to
April~2026), capturing the full 2025 AI adoption wave. Generalization
to non-open-source contexts or non-GitHub forges requires further
analysis; the Ecosyste.ms augmentation (Section~\ref{sec:completeness})
partially mitigates the GitHub bias.

\section{Conclusion and Future Work}
\label{sec:conclusion}

We presented a multi-layered detection framework for identifying AI
coding agent usage across the open-source ecosystem. Our taxonomy of
four trace types, centralized bot accounts, commit-message
signatures, distributed author-name patterns, and
configuration-file-only presence, demonstrates that no single
detection method captures the full extent of AI adoption: for Claude
Code, bot-account lookup alone undercounts by $30\times$. Applying
the framework to 180M+ repositories in three World of Code snapshots,
we identified twelve agents, documented the 2025 adoption wave at
ecosystem scale, and validated every detection cell with hand labels.
Adoption is bimodal (born-with-AI versus legacy integration), silent
agents dominate project-level presence, and configuration conventions
evolve rapidly, requiring continuous detector recalibration.

\paragraph{Data and code availability.}
The replication package releases the full detection pipeline as
parameterized scripts (Type~A--D shard scanners, the ClickHouse and
\texttt{agc2datFull} count drivers, and the aggregation step), the canonical
agent-pattern file that is the single source of truth for both the ClickHouse
and shard-scan paths, the 495-label hand-validation set with its
sampling/labeling/precision/recall scripts, the cross-method comparison code
and intermediate data for the AIDev analysis, and a spreadsheet mapping every
number in this paper to the script and output that produces it. The Claude
multi-method union and the per-cell precision counts are machine-verified to
reproduce. The package will be released on Zenodo upon publication.

To support replication, the labeled validation set, detection regex
patterns, and per-snapshot counts are included in this package.
Future work will extend the census to the forthcoming V2605 snapshot
(approximately 60 million additional repositories), cover agents not
yet included (Google Gemini Code Assist, OpenAI Codex/ChatGPT
beyond \texttt{AGENTS.md}), and develop standardized AI attribution
metadata for Git commits to reduce dependence on heuristic detection.

\appendix
\section{Appendix: Descriptive Tables}
\label{sec:appendix}

The following tables present naive descriptive before--after velocity for three agents. They use single-covariate selection on pre-velocity and do not control for project age, size, or contributor count. These descriptive tables are context only and do not control for confounders; they should not be read as effect sizes.

\begin{table}[h]
  \centering
  \small
  \begin{tabular}{lrrrrl}
    \toprule
    \textbf{Project} & \textbf{OH} & \textbf{Before} & \textbf{After}
      & \textbf{$\Delta$\%} & \textbf{Category} \\
    \midrule
    posix4e\_chroniclesync      & 370 &     0 & 1,678 & N/A       & Greenfield \\
    lizasaravia\_chronicle-sync &  45 &     3 &   310 & +10,233\% & Greenfield \\
    all-hands-ai\_oh-monitor    &  36 &     2 &   187 & +9,250\%  & Greenfield \\
    longisland\_multiplayer     &  86 &     6 &   335 & +5,483\%  & Near-green. \\
    kyu3q\_mini-erp-system      &  45 &     4 &   185 & +4,525\%  & Near-green. \\
    yadurshanm\_notes-app-next  &  25 &     6 &   105 & +1,650\%  & Near-green. \\
    jisaf\_location-map         &  33 &    26 &   129 & +396\%    & Legacy (+) \\
    zxy101\_mokuro-reader       &  34 &    49 &   203 & +314\%    & Legacy (+) \\
    all-hands-ai\_oh-aci        &  32 &   136 &   316 & +132\%    & Legacy (+) \\
    runtimerevolution\_revent   &  24 &    38 &    71 & +87\%     & Legacy (+) \\
    exciles\_openhands          & 299 &13,110 & 8,096 & $-$38\%   & Fork \\
    \bottomrule
  \end{tabular}
  \caption{Naive commit velocity before and after OpenHands adoption (6-month window). Descriptive only; includes greenfield projects.}
  \label{tab:impact_appendix}
\end{table}

\begin{table}[h]
\centering
\small
\begin{tabular}{lrrrrl}
\toprule
\textbf{Project} & \textbf{Aider} & \textbf{Before}
  & \textbf{After} & \textbf{$\Delta$\%} & \textbf{Category} \\
\midrule
int-dist-sys\_realtime-cli & 100 & 0 & 200 & N/A & Greenfield \\
jamespacileo\_prompt-manager & 100 & 0 & 200 & N/A & Greenfield \\
dredozubov\_advisor & 100 & 1 & 199 & +19,800\% & Near-green. \\
revelaction\_ical-git & 100 & 14 & 165 & +1,079\% & Near-green. \\
lesterpaintstheworld\_terminal & 95 & 65 & 135 & +108\% & Legacy (+) \\
morteng\_sf4 & 100 & 133 & 67 & $-$50\% & Legacy ($-$) \\
substratelabs\_rob-agi & 99 & 141 & 52 & $-$63\% & Legacy ($-$) \\
mbodiai\_mbpy & 100 & 158 & 42 & $-$73\% & Legacy ($-$) \\
\bottomrule
\end{tabular}
\caption{Naive commit velocity before and after Aider adoption (6-month window, sampled from top 20 authors, 100-commit-per-author cap). Descriptive only.}
\label{tab:aider_impact_appendix}
\end{table}

\paragraph{Event-study analysis.}
To assess the plausibility of the parallel-trends assumption required for Difference-in-Differences (DiD) analysis, we computed monthly commit counts within an 11-month window spanning six months before and five months after OpenHands adoption for the four legacy-eligible projects with sufficient pre-adoption history. Table~\ref{tab:event_study} summarizes the results.

All four projects exhibit a clear step-change immediately
following adoption (month $+1$), while pre-treatment
activity remains relatively flat or minimal. These four
projects are not the DiD sample (which uses Claude Code
legacy projects); they are shown here as descriptive
reference only. Aggregated across all four projects, total
commit volume increases from 151 commits in the
pre-adoption period to 752 commits post-adoption,
corresponding to a 4.98$\times$ increase. At the
individual project level, the immediate post-adoption
increase ranges from 2.0$\times$
(\texttt{all-hands-ai/openhands-aci}) to 103$\times$
(\texttt{runtimerevolution/revent-api}).

\begin{table}[h]
\centering
\small
\setlength{\tabcolsep}{4pt}
\begin{tabular}{lrrrrrr|rrrrr}
\toprule
\textbf{Project} & \multicolumn{6}{c|}{\textbf{Pre-adoption (months)}} & \multicolumn{5}{c}{\textbf{Post-adoption (months)}} \\
 & $-6$ & $-5$ & $-4$ & $-3$ & $-2$ & $-1$ & $+1$ & $+2$ & $+3$ & $+4$ & $+5$ \\
\midrule
all-hands-ai\_oh-aci       & 0 & 0 & 0 & 0 &  0 & 41 & 84 & 61 & 125 & 39 & 7 \\
jisaf\_location-map        & 0 & 0 & 0 & 0 &  0 & 20 & 129 & 0 &  0 &  0 & 0 \\
runtimerevolution\_revent  & 0 & 1 & 44 & 4 &  0 &  0 & 103 & 0 &  1 &  0 & 0 \\
zxy101\_mokuro-reader      & 2 & 1 &  0 & 1 & 32 &  5 &  71 & 110 & 22 &  0 & 0 \\
\midrule
\textbf{Total}             & 2 & 2 & 44 & 5 & 32 & 66 & 387 & 171 & 148 & 39 & 7 \\
\bottomrule
\end{tabular}
\caption{Monthly commit counts for four legacy OpenHands-adopting projects in the event-study window. The vertical divider separates pre- and post-adoption periods. Total pre-adoption commits: 151; total post-adoption commits: 752; overall increase: 4.98$\times$.}
\label{tab:event_study}
\end{table}

\newpage
\bibliographystyle{unsrt}
\bibliography{references}\end{document}